\documentstyle[12pt]{article}

\begin{document}

\makeatletter
\@addtoreset{equation}{section}
\def\theequation{\thesection.\arabic{equation}}
\makeatother

\vskip 0.5 truecm
\vskip 0.5 truecm

\begin{center}
{\large{\bf Linearity of quantum probability measure and Hardy's model
}}
\end{center}
\vskip .5 truecm
\centerline{\bf  Kazuo Fujikawa$^1$, C.H. Oh$^{2}$ and Chengjie Zhang$^{2}$}
\vskip .4 truecm
\centerline {\it $^1$ Mathematical Physics Laboratory,}
\centerline {\it RIKEN Nishina Center, Wako 351-0198, Japan}
\vspace{0.3cm}
\centerline {\it $^2$ Center for Quantum Technologies,}
\centerline {\it  National University of Singapore, Singapore 117543, Singapore}
\vskip 0.5 truecm

\begin{abstract}
We re-examine d=4 hidden-variables-models for a system of two spin-$1/2$ particles in view of the concrete model of Hardy, who analyzed the criterion of entanglement without referring to inequality.	The basis of our analysis is the linearity of the probability measure related to the Born probability interpretation, which excludes non-contextual hidden-variables models in $d\geq 3$. 
To be specific, we note the inconsistency of the non-contextual hidden-variables model in $d=4$ with the linearity of the quantum mechanical probability measure in the sense  $\langle\psi|{\bf a}\cdot {\bf \sigma}\otimes{\bf b}\cdot {\bf \sigma}|\psi\rangle+\langle\psi|{\bf a}\cdot {\bf \sigma}\otimes{\bf b}^{\prime}\cdot {\bf \sigma}|\psi\rangle=\langle\psi|{\bf a}\cdot {\bf \sigma}\otimes ({\bf b}+{\bf b}^{\prime})\cdot {\bf \sigma}|\psi\rangle$ for non-collinear ${\bf b}$ and ${\bf b}^{\prime}$.  
It is then shown that Hardy's model in $d=4$ does not lead to a unique mathematical expression in the demonstration of  the discrepancy of local realism (hidden-variables model) with entanglement and thus his proof is incomplete. We identify the origin of this non-uniqueness with the  non-uniqueness of translating quantum mechanical expressions into expressions in hidden-variables models, which results from the failure of the above linearity of the probability measure. 
In contrast, if the linearity of the probability measure is strictly imposed, which is tantamount to asking that the non-contextual hidden-variables model in $d=4$  gives the CHSH inequality $|\langle B\rangle|\leq 2$ uniquely, it is shown that the hidden-variables model can describe only separable quantum mechanical states; this conclusion is in perfect agreement with the so-called Gisin's theorem which states that $|\langle B\rangle|\leq 2$ implies separable states. 
\end{abstract}

%\maketitle

\section{Introduction}
The characterization of entanglement by inequalities is well known~\cite{bell,chsh, cs}. The conventional way of deriving such inequalities is to use a suitable  hidden-variables model as an auxiliary device to define correlations. On the other hand, Hardy~\cite{hardy1,hardy2} proposed the characterization of entanglement, which does not use inequalities, by employing EPR-type arguments with the minimum use of hidden-variables models. His analysis, in particular, the proposed gedanken experiment~\cite{hardy1}, is quite ingenious. Also, the generalization of his scheme to multi-party systems has been recently discussed~\cite{oh}.
Despite of those attractive features of his proposal, it is still disturbing that his scheme, which is intended as a measure of entanglement, completely fails for the maximally entangled case~\cite{hardy2}. 
We here re-examine his analysis by utilizing the more detailed information about the non-contextual hidden-variables models in $d=4$, in particular, the failure of the linearity of the quantum mechanical probability measure. We show that his model in $d=4$ does not lead to a unique conclusion in the demonstration of the inconsistency of local realism (non-contextual hidden-variables model) with quantum mechanical entanglement. 

We start with the definition of a non-contextual hidden-variables model~\footnote{We use the term "hidden-variables model" for the model in (1.1), and the adjective "non-contextual" is added when we emphasize other implicit constraints imposed on the model such as its validity for any $\psi$ with a uniform $P(\lambda)$.}
in $d=4=2\times 2$ dimensions of the Hilbert space~\cite{beltrametti}
\begin{eqnarray}
\langle\psi|{\bf a}\cdot {\bf \sigma}\otimes{\bf b}\cdot {\bf \sigma}|\psi\rangle=\int_{\Lambda} P(\lambda)d\lambda a(\psi, \lambda)b(\psi, \lambda),
\end{eqnarray}
where ${\bf a}$ and ${\bf b}$ are 3-dimensional unit vectors and ${\bf \sigma}$ stands for the Pauli matrix, and 
$a(\psi, \lambda)$ and $b(\psi, \lambda)$ are dichotomic variables assuming the eigenvalues $\pm 1$ of ${\bf a}\cdot {\bf \sigma}$ and ${\bf b}\cdot {\bf \sigma}$. This definition of the hidden-variables model agrees with
Eq. (2) of Bell~\cite{bell} and Eq. (3.5) of Clauser and Shimony~\cite{cs}.

The basic operational rule of the hidden-variables model is to translate a quantum mechanical
statement into the language of the hidden-variables model and then after the manipulations allowed in the hidden-variables model, translate the final result back into a statement of quantum mechanics which may be tested by experiments~\cite{bell}. A subtle aspect of this procedure is that the quantum mechanically equivalent  expressions are translated into in-equivalent expressions in  the hidden-variables model in general. In view of the absence of non-contextual hidden-variables models fully consistent with quantum mechanics in the Hilbert space with dimensions $d\geq 3$~\cite{gleason, bell2, kochen}, this ambiguity is inevitable and this makes the subjects related to hidden-variables models  subtle.
  Only when a statement is proved later purely in terms of the quantum mechanical language, as is the case of CHSH inequality\cite{werner, gisin}, the statement becomes solid.

To be more specific, we discuss the implications of the failure of the linearity condition of the probability measure~\cite{neumann} in the sense, 
\begin{eqnarray}
\langle\psi|{\bf a}\cdot {\bf \sigma}\otimes{\bf b}\cdot {\bf \sigma}|\psi\rangle+\langle\psi|{\bf a}\cdot {\bf \sigma}\otimes{\bf b}^{\prime}\cdot {\bf \sigma}|\psi\rangle=|{\bf b}+{\bf b}^{\prime}|\langle\psi|{\bf a}\cdot {\bf \sigma}\otimes \tilde{{\bf b}}\cdot {\bf \sigma}|\psi\rangle
\end{eqnarray}
in the hidden-variables representation (1.1), where 
$\tilde{{\bf b}}=({\bf b}+{\bf b}^{\prime})/|{\bf b}+{\bf b}^{\prime}|$    
for {\em non-collinear} ${\bf b}$ and ${\bf b}^{\prime}$. The operator
$|{\bf b}+{\bf b}^{\prime}|[{\bf a}\cdot {\bf \sigma}\otimes \tilde{{\bf b}}\cdot {\bf \sigma}]$ corresponds to the spectral decomposition of ${\bf a}\cdot {\bf \sigma}\otimes ({\bf b}+{\bf b}^{\prime})\cdot {\bf \sigma}$. Both hands of (1.2) are separately evaluated by the formula (1.1), but the results do not agree in general. 
This linearity condition of the probability measure becomes crucial in our analysis of Hardy's model.     

We here  briefly mention the linearity of the probability measure and non-contextuality in a general context. Gleason~\cite{gleason} analyzed the
possible  probability measure $v$ in quantum mechanics which assigns non-negative values $v(P_{k})$ to any set of complete orthogonal projection operators
\begin{eqnarray}
\sum_{k}P_{k}=1
\end{eqnarray}
by preserving the {\em linearity condition} $v(\sum_{k}P_{k})=\sum_{k}v(P_{k})=1$. He concludes that such a measure is inevitably given by a trace representation $v(P_{k})=Tr (\rho P_{k})$ with a suitable trace-class operator $\rho$ for the dimensions of the Hilbert space $d\geq 3$. This shows that such a measure is not dispersion-free and that the linearity is extended to any hermitian operators $Tr \rho (A+B)=Tr \rho A + Tr \rho B$. The linearity condition (1.2) is related to this last property. Here {\em dispersion-free} means the assignment of eigenvalues $1$ or $0$ to $v(P_{k})$ for any projection operators $P_{k}$. On the other hand, Kochen and Specker~\cite{kochen} explicitly demonstrated that the dispersion-free and non-contextual measure such as suggested by hidden-variables models, $\sum_{k}v(P_{k})=1$ with $v(P_{k})=1 \ {\rm or}\ 0$ for any orthogonal $P_{k}$ appearing in the sum $\sum_{k}P_{k}=1$,  gives rise to a contradiction in $d=3$, although their actual analysis is performed using the square of angular momentum operators. 

The superposition principle, which is responsible for entanglement, and the linearity of the probability measure we discuss together form the basis of quantum mechanics. The superposition principle arises from the fact that the Schr\"{o}dinger equation is linear, while the linearity of the probability measure we discuss is fundamental to the Born probability interpretation~\cite{neumann}. The superposition principle deals with probability amplitudes while the linearity of the probability measure deals with probability itself. The existence or absence of hidden-variables models is more directly related to the linearity of the probability measure; the absence of non-contextual hidden-variables models in $d\geq 3$ is concluded from the linearity of the probability measure without referring to the superposition principle~\cite{gleason, kochen}. We analyze Hardy's concrete model  in $d=4$ with emphasis on this fundamental linearity condition. 
 
\section{Hardy's model}
We here recapitulate the analysis of Hardy~\cite{hardy1,hardy2}.
Hardy chooses the projection operators~\cite{hardy2}
\begin{eqnarray}
\hat{U}_{i}=|u_{i}\rangle\langle u_{i}|,\ \
\hat{D}_{i}=|d_{i}\rangle\langle d_{i}|,
\end{eqnarray}
with $i=1,2$, and 
\begin{eqnarray}
|u_{i}\rangle&=&\frac{1}{\sqrt{\alpha+\beta}}[\beta^{1/2}|+\rangle_{i}+\alpha^{1/2}|-\rangle_{i}],\nonumber\\
|d_{i}\rangle&=&\frac{1}{\sqrt{\alpha^{3}+\beta^{3}}}[\beta^{3/2}|+\rangle_{i}-\alpha^{3/2}|-\rangle_{i}]
\end{eqnarray}
for the entangled state
\begin{eqnarray}
|\psi\rangle=\alpha|+\rangle_{1}|+\rangle_{2}-\beta|-\rangle_{1}|-\rangle_{2}
\end{eqnarray}
with $\alpha^{2}+\beta^{2}=1$.
We work with real and non-negative $\alpha$ and $\beta$, for simplicity, but more general cases can be treated similarly.

He then obtains the relations (note that $D_{1}U_{2}D_{1}=D_{1}U_{2}$, for example)
\begin{eqnarray}
\frac{\langle\psi|D_{1}U_{2}D_{1}|\psi\rangle}{\langle\psi|D_{1}|\psi\rangle}&=&1,\\
\frac{\langle\psi|D_{2}U_{1}D_{2}|\psi\rangle}{\langle\psi|D_{2}|\psi\rangle}&=&1,\\
\frac{\langle\psi|D_{1}D_{2}D_{1}|\psi\rangle}{\langle\psi|D_{1}|\psi\rangle}&=&1-\frac{\alpha\beta}{(1-\alpha\beta)},
\\
\langle\psi|D_{1}|\psi\rangle=\langle\psi|D_{2}|\psi\rangle
&=&\frac{\alpha^{2}\beta^{2}}{1-\alpha\beta},\\
\langle\psi|U_{1}|\psi\rangle
=\langle\psi|U_{2}|\psi\rangle
&=&\alpha\beta,
\end{eqnarray}
and 
\begin{eqnarray}
\langle\psi|U_{1}U_{2}|\psi\rangle=0,
\end{eqnarray}
with $0< \alpha\beta\leq 1/2$.

Hardy then argues that~\cite{hardy2}:\\
i)The measured value of $\langle D_{1}D_{2}\rangle\neq 0$ in (2.6) for $0< \alpha\beta< 1/2$ implies 
$D_{1}(\psi, \lambda)=D_{2}(\psi, \lambda)=1$ for {\em some} $\lambda \in \Lambda$, which is based on the  assumption that the hidden-variables model
\begin{eqnarray}
\langle\psi| D_{1}D_{2}|\psi\rangle=\int_{\Lambda} P(\lambda)d\lambda D_{1}(\psi, \lambda)D_{2}(\psi, \lambda),
\end{eqnarray}
is valid for this combination.\\
ii)The assumption  of the validity of the conditional probability
\begin{eqnarray}
\frac{\langle\psi|D_{2}U_{1}D_{2}|\psi\rangle}{\langle\psi|D_{2}|\psi\rangle}&=&\frac{\int_{\Lambda} P(\lambda)d\lambda U_{1}(\psi, \lambda)D_{2}(\psi, \lambda)
}{\int_{\Lambda} P(\lambda)d\lambda D_{2}(\psi, \lambda)}\nonumber\\
&=&1,
\end{eqnarray}
implies  $U_{1}(\psi, \lambda)=1$ for all $\lambda$ for which $D_{2}(\psi, \lambda)=1$.\\
iii)Similarly, the assumption  of the validity of  the conditional probability
\begin{eqnarray}
\frac{\langle\psi|D_{1}U_{2}D_{1}|\psi\rangle}{\langle\psi|D_{1}|\psi\rangle}&=&\frac{\int_{\Lambda} P(\lambda)d\lambda U_{2}(\psi, \lambda)D_{1}(\psi, \lambda)
}{\int_{\Lambda} P(\lambda)d\lambda D_{1}(\psi, \lambda)}\nonumber\\&=&1
,
\end{eqnarray}
implies  $U_{2}(\psi, \lambda)=1$ for all $\lambda$ for which $D_{1}(\psi, \lambda)=1$.\\
iv) The assumption
\begin{eqnarray}
\langle\psi| U_{1}U_{2}|\psi\rangle=\int_{\Lambda} P(\lambda)d\lambda U_{1}(\psi, \lambda)U_{2}(\psi, \lambda),
\end{eqnarray}
then implies that $\langle\psi| U_{1}U_{2}|\psi\rangle\neq 0$, but this contradicts the prediction of quantum mechanics $\langle\psi| U_{1}U_{2}|\psi\rangle= 0$ in (2.9). Hardy suggests that this provides a  test of nonlocality for the entangled state without inequalities~\cite{hardy2}, namely, local realism (hidden-variables model) cannot explain  entanglement.  It is significant that his reasoning
does not work for the maximally entangled state with $\alpha\beta=1/2$.

Instead of the above analysis of Hardy, one may equally argue that the relation in iv), $\langle\psi| U_{1}U_{2}|\psi\rangle=0$, is measured and thus
\begin{eqnarray}
\int_{\Lambda} P(\lambda)d\lambda U_{1}(\psi, \lambda)U_{2}(\psi, \lambda)=0,
\end{eqnarray}
is the prediction of  the hidden-variables model. If relations ii) and iii) are correct, the relation i) becomes inconsistent for $0<\alpha\beta<1/2$, since 
\begin{eqnarray}
\langle\psi| D_{1}D_{2}|\psi\rangle=\int_{\Lambda} P(\lambda)d\lambda D_{1}(\psi, \lambda)D_{2}(\psi, \lambda)
=0
\end{eqnarray}
is concluded from those relations, contrary to (2.6); in terms of hidden-variables
space, (2.11) implies that the domain with $D_{2}(\psi, \lambda)=1$ is included in the domain with $U_{1}(\psi, \lambda)=1$ (which is consistent with $\langle\psi|U_{1}|\psi\rangle>\langle\psi|D_{2}|\psi\rangle$), and similarly (2.12) implies that the domain with $D_{1}(\psi, \lambda)=1$ is included in the domain with $U_{2}(\psi, \lambda)=1$ (which is consistent with $\langle\psi|U_{2}|\psi\rangle>\langle\psi|D_{1}|\psi\rangle$). But the relation (2.14) shows that there is no common domain for $U_{1}(\psi, \lambda)=1$ and $U_{2}(\psi, \lambda)=1$, and thus no common domain for $D_{1}(\psi, \lambda)=1$ and $D_{2}(\psi, \lambda)=1$. One thus arrives at (2.15) and a contradiction with quantum mechanical predictions.

Obviously, Hardy's model tests the consistency of the  hidden-variables model defined in (1.1) with quantum mechanics, and not all the relations i)-iv) are simultaneously consistent with quantum mechanics.
For the given state $\psi$ for $0<\alpha\beta<1/2$, the quantum mechanical predictions (2.6) and (2.9) 
are uniquely defined. But we cannot decide which of 
\begin{eqnarray}
&&\int_{\Lambda} P(\lambda)d\lambda D_{1}(\psi, \lambda)D_{2}(\psi, \lambda)\neq 0,
\nonumber\\
&&\int_{\Lambda} P(\lambda)d\lambda U_{1}(\psi, \lambda)U_{2}(\psi, \lambda)\neq 0,
\end{eqnarray}
or 
\begin{eqnarray}
&&\int_{\Lambda} P(\lambda)d\lambda U_{1}(\psi, \lambda)U_{2}(\psi, \lambda)= 0
,\nonumber\\
&&\int_{\Lambda} P(\lambda)d\lambda D_{1}(\psi, \lambda)D_{2}(\psi, \lambda)= 0
\end{eqnarray}
is the {\em prediction} of the hidden-variables model. We attribute this lack of uniqueness to the lack of uniqueness in translating the quantum mechanical expressions into the language of the hidden-variables model.

To analyze this issue, we examine the combination
\begin{eqnarray}
U_{i}+D_{i}&=&|u_{i}\rangle\langle u_{i}|+|d_{i}\rangle\langle d_{i}|.
\end{eqnarray}
If one sets $\alpha=\beta$, $U_{i}$ and $D_{i}$ become collinear and $U_{i}D_{i}=0$. For the generic case $\alpha\neq \beta$, we  perform the spectral decomposition
\begin{eqnarray}
U_{i}+D_{i}=\mu_{1}P^{(1)}_{i}+\mu_{2}P^{(2)}_{i},
\end{eqnarray}
with the orthogonal projection operators
\begin{eqnarray}
P^{(1)}_{i}+P^{(2)}_{i}=1, \ \ \ P^{(1)}_{i}P^{(2)}_{i}=0,
\end{eqnarray}
and two {\em positive} eigenvalues which satisfy $\mu_{1}+\mu_{2}=2$.
The only integral eigenvalues allowed are 
\begin{eqnarray}
&&\mu_{1}=\mu_{2}=1\ \ {\rm for}\ \  \alpha=\beta (\neq 0),\nonumber\\
&&\mu_{1}=2,\ \mu_{2}=0\ \ {\rm for}\ \  \alpha\neq 0,\ \beta=0,\nonumber\\
&&\mu_{1}=0,\ \mu_{2}=2\ \ {\rm for}\ \  \alpha=0,\ \beta\neq 0.
\end{eqnarray}
Thus we have the  relation corresponding to the no-go argument of von Neumann~\cite{neumann}, 
\begin{eqnarray}
U_{i}(\lambda)+D_{i}(\lambda)\neq \mu_{1}P^{(1)}_{i}(\lambda)+\mu_{2}P^{(2)}_{i}(\lambda),
\end{eqnarray}
or
\begin{eqnarray}
U_{i}(\lambda)+D_{i}(\lambda)\neq (U_{i}+D_{i})(\lambda),
\end{eqnarray}
for the range of parameters  $0<\alpha\beta<1/2$ for {\em any} hidden-variables $\lambda$ when one assumes dispersion-free representations. Bell argued~\cite{bell2} that we should not use the relation (2,23) itself but only the quantities integrated over hidden variables
$\lambda$; we expect that the inequality in (2.23) is converted to an equality 
after the integration over hidden variables in the formula such as (1.1) if 
the hidden-variables model reproduces the result of quantum mechanics. This condition is indeed satisfied by the $d=2$ models of Bell~\cite{bell2} and Kochen-Specker~\cite{kochen}, but not by the generic $d=4$ model in (1.1)~\cite{gleason, kochen, beltrametti} as is further explained in Section 3.

By exploiting this lack of linearity in the probability measure, which is characteristic to a system with non-commuting operators, we suggest one possible resolution of the above (2.16) and (2.17) by adopting 
\begin{eqnarray}
\langle \psi|U_{1}U_{2}|\psi\rangle&=&\langle \psi|(U_{1}+D_{1})U_{2}|\psi\rangle-\langle\psi| D_{1}U_{2}|\psi\rangle\nonumber\\
&\equiv&\int_{\Lambda}d\lambda P(\lambda)(U_{1}+D_{1})(\psi, \lambda)U_{2}(\psi, \lambda)\nonumber\\
&&-\int_{\Lambda}d\lambda P(\lambda)D_{1}(\psi, \lambda)U_{2}(\psi, \lambda)
\nonumber\\
&=&0,
\end{eqnarray}
as the {\em definition} of the hidden-variables representation of $\langle \psi|U_{1}U_{2}|\psi\rangle$ and $\langle \psi|(U_{1}+D_{1})U_{2}|\psi\rangle-\langle\psi| D_{1}U_{2}|\psi\rangle$, instead of the standard (2.13). Namely, we translate one of the equivalent quantum mechanical expressions into a hidden-variables representation. 
A motivation for using this lack of the linearity relation comes from the analysis of CHSH inequality. If one starts with the quantum mechanical operator $B$ introduced by Cirel'son~\cite{cirel'son}
\begin{eqnarray}
B={\bf a}\cdot {\bf \sigma}\otimes ({\bf b}+{\bf b}^{\prime})\cdot {\bf \sigma} +{\bf a}^{\prime}\cdot{\bf \sigma}\otimes ({\bf b}-{\bf b}^{\prime})\cdot{\bf \sigma},
\end{eqnarray}
where ${\bf a},\ {\bf a}^{\prime},\ {\bf b},\ {\bf b}^{\prime}$ are 3-dimensional unit vectors, one has the quantum mechanically equivalent  relations
\begin{eqnarray}
\langle B\rangle&=&\langle {\bf a}\cdot {\bf \sigma}\otimes ({\bf b}+{\bf b}^{\prime})\cdot {\bf \sigma}\rangle +\langle {\bf a}^{\prime}\cdot{\bf \sigma}\otimes ({\bf b}-{\bf b}^{\prime})\cdot{\bf \sigma}\rangle\\
&=&\langle {\bf a}\cdot {\bf \sigma}\otimes {\bf b}\cdot {\bf \sigma}\rangle +
\langle{\bf a}\cdot {\bf \sigma}\otimes {\bf b}^{\prime}\cdot {\bf \sigma}\rangle +\langle {\bf a}^{\prime}\cdot{\bf \sigma}\otimes {\bf b}\cdot{\bf \sigma}\rangle-\langle {\bf a}^{\prime}\cdot{\bf \sigma}\otimes {\bf b}^{\prime}\cdot{\bf \sigma}\rangle \nonumber
\end{eqnarray}
If one moves to the hidden-variables representation from the last expression, one obtains the standard CHSH relation $|\langle B\rangle|\leq 2$~\cite{chsh}, while if one moves to the hidden-variables representation from the first expression one obtains $|\langle B\rangle|\leq 2\sqrt{2}$~\cite{fujikawa}. Namely, the failure of the linearity in the probability measure leads to different physical predictions of hidden-variables models. 

Coming back to the problem at hand, (2.16) then becomes (qualitatively) consistent with
quantum mechanics since 
\begin{eqnarray}
&& \int_{\Lambda} P(\lambda)d\lambda U_{1}(\psi, \lambda)U_{2}(\psi, \lambda)
\nonumber\\
&\neq&\int_{\Lambda}d\lambda P(\lambda)(U_{1}+D_{1})(\psi, \lambda)U_{2}(\psi, \lambda)-\int_{\Lambda}d\lambda P(\lambda)D_{1}(\psi, \lambda)U_{2}(\psi, \lambda)
\nonumber\\
&=&0
\end{eqnarray}
for $0<\alpha\beta<1/2$, and the issue (2.17) does not arise; the inequality in (2.27), which has the same structure as (1.2) for the non-contextual hidden-variables model (1.1), is shown later in Section 3. The choice (2.24) is rather arbitrary  but it has an advantage since 
\begin{eqnarray}
&& \int_{\Lambda} P(\lambda)d\lambda U_{1}(\psi, \lambda)U_{2}(\psi, \lambda)
\nonumber\\
&=&\int_{\Lambda}d\lambda P(\lambda)(U_{1}+D_{1})(\psi, \lambda)U_{2}(\psi, \lambda)-\int_{\Lambda}d\lambda P(\lambda)D_{1}(\psi, \lambda)U_{2}(\psi, \lambda)
\nonumber\\
&=&0
\end{eqnarray}
for $\alpha\beta=1/2$, for which $U_{1}$ and $D_{1}$ are orthogonal 
$U_{1}D_{1}=0$ and commuting and thus the linearity is restored, $(U_{1}+D_{1})(\psi, \lambda)=U_{1}(\psi, \lambda)+D_{1}(\psi, \lambda)$. One thus obtains $\langle \psi|U_{1}U_{2}|\psi\rangle=\int_{\Lambda} P(\lambda)d\lambda U_{1}(\psi, \lambda)U_{2}(\psi, \lambda)$ and in fact, for $\alpha\beta=1/2$, (2.17) is consistent with the quantum mechanical predictions (2.6) and (2.9). One can thus treat all the range of the parameter
$0<\alpha\beta\leq 1/2$ uniformly including the maximally entangled state without contradicting quantum mechanics in any obvious way, provided that one explicitly admits the failure of linearity in the probability measure in the hidden-variables representation (1.1) in $d=4$.

This resolution of (2.16) and (2.17) illustrates that Hardy's model
tests the consistency of the hidden-variables model with the linearity of the probability measure, i.e., tests the uniqueness of various expressions in the hidden-variables model such as $\langle \psi|U_{1}U_{2}|\psi\rangle=\int_{\Lambda} P(\lambda)d\lambda U_{1}(\psi, \lambda)U_{2}(\psi, \lambda)$, in addition to entanglement. This lack of uniqueness ultimately implies the failure of the hidden-variables model in describing quantum mechanics without referring to entanglement, in addition to the failure of the hidden-variables model in describing  entanglement, to be consistent with the absence of non-contextual hidden-variables models in $d=4$~\cite{beltrametti, gleason, kochen}. 

Finally, we comment on the specific properties of Hardy's model.
Since the model of Hardy is entangled for $0<\alpha\beta$, any inconsistency of the hidden-variables model with quantum mechanical predictions might formally be taken as an indication of the 
inability to describe entanglement by the hidden-variables model. But this argument needs to be scrutinized. For the maximally entangled case $\alpha\beta=1/2$, the hidden-variables model and quantum mechanics give consistent predictions. 
It is interesting to see how the consistency of quantum mechanics and the hidden-variables model can be realized for the maximally entangled case  $\alpha\beta=1/2$ in Hardy's model.
We have the dispersion-free representations 
\begin{eqnarray}
&&D_{1}(\psi,\lambda)+U_{1}(\psi,\lambda)=1,\ \
D_{1}(\psi,\lambda)U_{1}(\psi,\lambda)=0,\nonumber\\
&&D_{2}(\psi,\lambda)+U_{2}(\psi,\lambda)=1,\ \
D_{2}(\psi,\lambda)U_{2}(\psi,\lambda)=0,
\end{eqnarray}
which are the standard hidden-variables representations of orthogonal complete $d=2$ projectors,
and the supplementary conditions
\begin{eqnarray}
&&D_{1}(\psi,\lambda)=U_{2}(\psi,\lambda),\ \
D_{2}(\psi,\lambda)=U_{1}(\psi,\lambda),
\end{eqnarray}
which may be adopted for the present special choice of operators and the state $\psi$. Note that these conditions are chosen consistently with $\langle \psi|U_{1}|\psi\rangle=\langle \psi|D_{1}|\psi\rangle=\langle \psi|U_{2}|\psi\rangle=\langle \psi|D_{2}|\psi\rangle=1/2$.
Then we have from (2.29) and (2.30),
\begin{eqnarray}
D_{1}(\psi,\lambda)D_{2}(\psi,\lambda)=0,\ \
U_{1}(\psi,\lambda)U_{2}(\psi,\lambda)=0,
\end{eqnarray}
and 
\begin{eqnarray}
&&U_{2}(\psi,\lambda)D_{1}(\psi,\lambda)=D_{1}(\psi,\lambda),\nonumber\\
&&U_{1}(\psi,\lambda)D_{2}(\psi,\lambda)=D_{2}(\psi,\lambda),
\end{eqnarray}
which reproduce the results of quantum mechanics in (2.4)-(2.9) after integration over hidden-variables in the hidden-variables representation for the special set of operators 
in (2.1), $\hat{U}_{i}=|u_{i}\rangle\langle u_{i}|$ and 
$\hat{D}_{i}=|d_{i}\rangle\langle d_{i}|$ with $\langle u_{i}|d_{i}\rangle=0$, and the state
\begin{eqnarray}
\psi=\frac{1}{\sqrt{2}}[|u_{1}\rangle|d_{2}\rangle+|d_{1}\rangle|u_{2}\rangle].
\end{eqnarray}
A salient feature of Hardy's model is that both of operators and  the state are simultaneously controlled by the same parameters $\alpha$ and $\beta$ and a very limited set of operators are considered; if one allows a more general class  of operators, a consistent hidden-variables model does not exist.

\section{Linearity of probability measure}
The linearity of the probability measure  is fundamental  to quantum mechanics from the days of von Neumann~\cite{neumann}.  By analyzing the linearity, Gleason~\cite{gleason} showed that quantum theory  in the Hilbert space with dimensions $d\geq 3$ is expressed by the trace formula thus excluding the dispersion-free non-contextual hidden-variables models in $d\geq 3$. In $d=2$ we know that non-contextual hidden-variables models consistent with linearity condition are given by Bell~\cite{bell2} and Kochen-Specker~\cite{kochen}. Busch~\cite{busch} has shown that the analysis of Gleason is generalized to $d=2$ if one extends projection operators to positive operator valued measures. In this general setting of POVMs, all the sensible non-contextual hidden-variables models in any dimensions of Hilbert space are excluded. The conditional 
measurement is the first example where the necessity of POVMs was recognized~\cite{umegaki}. It has been recently pointed out~\cite{fujikawa2} that the well-known examples of non-contextual hidden-variables models in $d=2$ by Bell~\cite{bell2} and Kochen-Specker~\cite{kochen} have certain difficulties when applied to conditional measurements. 

It may be interesting to examine what happens if one imposes linearity of the probability measure  strictly on the non-contextual hidden-variables model in $d=4$ as defined in (1.1), instead of the accidental recovery of linearity in (2.28). We here briefly report the essence of this analysis~\cite{fujikawa}.
To analyze the linearity condition, we start with relations in quantum mechanics,
\begin{eqnarray}
&&\langle (\hat{a}_{1}+\hat{b}_{1})\otimes \hat{1} \rangle_{\psi}=\langle \hat{a}_{1}\otimes \hat{1} \rangle_{\psi}+\langle \hat{b}_{1}\otimes \hat{1} \rangle_{\psi}, \\
&&\langle (\hat{a}_{1}+\hat{b}_{1})\otimes \hat{a}_{2} \rangle_{\psi}=\langle \hat{a}_{1}\otimes \hat{a}_{2} \rangle_{\psi}+\langle \hat{b}_{1}\otimes \hat{a}_{2} \rangle_{\psi}, 
\end{eqnarray}
where $\hat{a}_{1}$, $\hat{b}_{1}$ and $\hat{a}_{2}$ are $d=2$ projection operators; for example, $\hat{a}_{1}=(1+{\bf a}_{1}\cdot {\bf \sigma})/2$ in the notation of (1.1). One can equally work with the representation (1.1) and (1.2), but following Hardy  we analyze projection operators. In particular, we choose $\hat{a}_{1}$ and $\hat{b}_{1}$ to be {\em non-collinear}. All the operators appearing in the relations (3.1) and (3.2) are well-specified in the hidden-variables model, and for the operator $\hat{a}_{1}+\hat{b}_{1}$ we perform a spectral decomposition
\begin{eqnarray}
\hat{a}_{1}+\hat{b}_{1}=\mu_{1}\hat{P}_{1}+\mu_{2}\hat{P}_{2},
\end{eqnarray}
and thus $(\hat{a}_{1}+\hat{b}_{1})\otimes \hat{a}_{2}=\mu_{1}\hat{P}_{1}\otimes \hat{a}_{2}+\mu_{2}\hat{P}_{2}\otimes \hat{a}_{2}$ with $\hat{P}_{1}+\hat{P}_{2}=1$ and $\hat{P}_{1}\hat{P}_{2}=0$. The left-hand side of (3.2) is then 
translated into the language of the hidden-variables model using a sum of two orthogonal projection operators
$\mu_{1}\langle \hat{P}_{1}\otimes \hat{a}_{2}\rangle_{\psi}+ \mu_{2}\langle \hat{P}_{2}\otimes \hat{a}_{2}\rangle_{\psi}$, 
to be consistent with the criterion of Bell~\cite{bell2}. 

The  relations corresponding to (3.1) and (3.2) are written in the framework of the non-contextual hidden-variables model in (1.1) as
\begin{eqnarray}
&&\int_{\Lambda}d\lambda P(\lambda)[(a_{1}+b_{1})(\psi, \lambda)- a_{1}(\psi, \lambda) - b_{1}(\psi, \lambda)]=0, \\
&&\int_{\Lambda}d\lambda P(\lambda)a_{2}(\psi, \lambda)[(a_{1}+b_{1})(\psi, \lambda)- a_{1}(\psi, \lambda) - b_{1}(\psi, \lambda)]=0,
\end{eqnarray}
where $a_{1}(\psi, \lambda),\ \ b_{1}(\psi, \lambda)$ and $a_{2}(\psi, \lambda)$
assume $1$ or $0$. The combination $(a_{1}+b_{1})(\psi, \lambda)$ is defined by
using the spectral decomposition $(a_{1}+b_{1})(\psi, \lambda)=\mu_{1}P_{1}(\psi, \lambda)+\mu_{2}P_{2}(\psi, \lambda)$.
The relations (3.4) and (3.5) should hold for all $\psi$ and for all choices of the operators $a$ and $b$ with a uniform  weight factor $P(\lambda)$. This 
is the definition of {\em non-contextual} hidden-variables models~\cite{bell, chsh, cs, beltrametti}.
\\

The analysis is most transparent if one adopts the parameterization suggested in the original paper of Bell~\cite{bell}
\begin{eqnarray}
P(\lambda)d\lambda=P(\lambda_{1},\lambda_{2})d\lambda_{1}d\lambda_{2}
\end{eqnarray}
and 
\begin{eqnarray}
a_{1}(\psi, \lambda_{1}),\ \ b_{1}(\psi, \lambda_{1}),\ \
a_{2}(\psi, \lambda_{2}),\ \ b_{2}(\psi, \lambda_{2}).
\end{eqnarray}
Namely, two systems are described by each hidden variables but still we maintain symmetry between two parties~\footnote{This parameterization in (3.6) and (3.7) is more general than the conventional one, since one can reproduce the conventional parameterization by choosing $P(\lambda_{1},\lambda_{2})=P(\lambda_{1})\delta(\lambda_{1}-\lambda_{2})$.}. 
In this choice, the relations (3.4) and (3.5) become
\begin{eqnarray}
&&\int_{\Lambda}P(\lambda_{1},\lambda_{2})d\lambda_{1}d\lambda_{2}[(a_{1}+b_{1})(\psi, \lambda_{1})- a_{1}(\psi, \lambda_{1}) - b_{1}(\psi, \lambda_{1})]=0, \\
&&\int_{\Lambda}P(\lambda_{1},\lambda_{2})d\lambda_{1}d\lambda_{2}a_{2}(\psi, \lambda_{2})[(a_{1}+b_{1})(\psi, \lambda_{1})- a_{1}(\psi, \lambda) - b_{1}(\psi, \lambda_{1})]=0.\nonumber\\
\end{eqnarray}
These relations may be regarded as defining a $d=2$ hidden-variables model  defined by $a_{1}(\psi, \lambda_{1})$ and $b_{1}(\psi, \lambda_{1})$  with the weight factor for the model given by 
\begin{eqnarray}
\frac{\int_{\Lambda}P(\lambda_{1},\lambda_{2})d\lambda_{2}}{\int_{\Lambda}P(\lambda_{1},\lambda_{2})d\lambda_{1}d\lambda_{2}}, \ \ {\rm and}\ \ 
\frac{\int_{\Lambda}P(\lambda_{1},\lambda_{2})a_{2}(\psi, \lambda_{2})d\lambda_{2}}{\int_{\Lambda}P(\lambda_{1},\lambda_{2})a_{2}(\psi, \lambda_{2})d\lambda_{1}d\lambda_{2}},
\end{eqnarray}
where the first relation implies a non-contextual $d=2$ model. 

Since (3.8) and (3.9) hold for any $\psi$, one may choose a separable state
$\psi=\psi_{1}\psi_{2}$.
 By choosing the state $\psi_{2}$ suitably for the fixed $a_{2}$, one can change
$a_{2}(\psi, \lambda_{2})=a_{2}(\psi_{2}, \lambda_{2})$ from unity for all $\lambda_{2}$ to zero for all $\lambda_{2}$ (namely, by changing the spin state $\psi_{2}$ which is parallel to $\hat{a}_{2}$ to the state which is anti-parallel to $\hat{a}_{2}$). If one assumes that the weight factor (3.10) is uniquely specified by the representations of $a_{1}(\psi_{1}, \lambda_{1})$ and $b_{1}(\psi_{1}, \lambda_{1})$, which is the case for the known $d=2$ models of Bell~\cite{bell2} and Kochen-Specker~\cite{kochen}, one then concludes 
\begin{eqnarray}
P(\lambda_{1},\lambda_{2})=P_{1}(\lambda_{1})P_{2}(\lambda_{2})
\end{eqnarray}
where a symmetry consideration may imply $P_{1}=P_{2}$. For the choice of (3.11), the two alternative expressions in (3.10) agree with each other. 
\\

We thus conclude that the non-contextual hidden-variables model in $d=4$ is written in the form 
\begin{eqnarray}
\langle\psi| \hat{a}_{1}\otimes\hat{b}_{2}|\psi\rangle&=& \int P_{1}(\lambda_{1})d\lambda_{1}a_{1}(\psi, \lambda_{1})
\int P_{2}(\lambda_{2})d\lambda_{2}b_{2}(\psi, \lambda_{2}).
\end{eqnarray}
 This representation implies 
\begin{eqnarray}
\langle\psi| \hat{a}_{1}\otimes\hat{b}_{2}|\psi\rangle&=& \langle\psi| \hat{a}_{1}\otimes 1|\psi\rangle\langle\psi|1\otimes\hat{b}_{2}|\psi\rangle
\end{eqnarray}
for any choice of $\hat{a}_{1}$ and $\hat{b}_{2}$, which clearly shows that the pure state $\psi$ in $d=4$ is separable. The right-hand side of (3.12) is then consistently represented by 
 a  factored product of 
$d=2$ Bell~\cite{bell2} or Kochen-Specker~\cite{kochen} or similar hidden-variables models which satisfies the linearity of the probability measure. Thus the model (1.1) is expressed by a product of dispersion-free representations in $d=2$ but it becomes trivial in the proper sense of the hidden-variables model in $d=4$. This conclusion is consistent with Gleason's theorem which excludes  fully $d=4$ non-contextual hidden-variables models. 

The $d=4$ non-contextual hidden-variables model in (1.1), when the linearity conditions (3.1) and (3.2) are imposed, can represent only a separable quantum mechanical state~\cite{fujikawa} and thus entanglement is completely missing.
This fact may  be taken as an example of the inconsistency of local realism or hidden-variables models with the quantum mechanical entanglement in the sense of Hardy without referring to inequalities. 

As for CHSH inequality,
using the operator $B$ in (2.25) introduced by Cirel'son~\cite{cirel'son}, one can readily derive 
$|\langle\psi|B|\psi\rangle|\leq 2$ for the separable state (3.12). 
The CHSH inequality was originally derived from the hidden-variables model without asking its consistency with quantum mechanics. Later, it was shown that CHSH inequality is the necessary and sufficient condition of the separability of  pure quantum mechanical states~\cite{werner, gisin}; formulated in this manner, CHSH inequality is valid without referring to hidden-variables models but its direct connection with local realism is lost. To be more explicit, $|\langle\psi|B|\psi\rangle|\leq 2$ implies the separable state $|\psi\rangle$~\cite{werner, gisin}. This fact is perfectly consistent with our derivation of a factored product of two $d=2$ hidden-variables models in (3.12), which is equivalent to a pure  separable state in $d=4$; we asked that the hidden-variables model gives CHSH inequalilty $|\langle\psi|B|\psi\rangle|\leq 2$ uniquely and we found that the hidden-variables model can describe only the separable state.
It is significant that CHSH inequality itself is consistent with the principles of quantum mechanics, in the sense that  separable states are consistent with  the principles of quantum mechanics.

The representation of the hidden-variables model (3.12), which satisfies the linearity of the probability measure but now equivalent to separable states, makes the model of Hardy trivial; the separable state is realized for $\alpha=0$ or $\beta=0$, but the model defined by (2.1)- (2.3) then gives rise to trivial results
\begin{eqnarray}
\langle U_{1}\rangle=\langle U_{2}\rangle=\langle D_{1}\rangle=\langle D_{2}\rangle=0,
\end{eqnarray}
and all the correlations are vanishing.
Only the case with $\alpha\beta=1/2$ as in (2.29), (2.30) and (2.33) gives rise to a consistent non-trivial hidden-variables representaion for the very limited set  of operators, and the physical predictions are $\langle U_{1}U_{2}\rangle=\langle D_{1}D_{2}\rangle=0$.

In passing, we mention the GHZ model in $d=8$~\cite{ghz}. We now analyze the hidden-variables model in analogy with (3.6),
\begin{eqnarray}
\langle\psi| \hat{a}_{1}\otimes\hat{b}_{2}\otimes\hat{c}_{3}|\psi\rangle&=& \int P(\lambda_{1},\lambda_{2},\lambda_{3})d\lambda_{1}d\lambda_{2}d\lambda_{3}a_{1}(\psi, \lambda_{1})b_{2}(\psi, \lambda_{2})c_{3}(\psi, \lambda_{3})
\end{eqnarray}
and impose the linearity condition so that we can apply the formula to the quantum 
mechanical state $\psi$. The final outcome is that the formula is written as a factored product of three $d=2$ non-contextual hidden-variables models, which is equivalent to a product of three quantum mechanical $d=2$ states. Unlike the prediction of the original entangled GHZ state which excludes local realism, our hidden-variables model is consistent with quantum mechanics and identical to a completely separable state; as a result it cannot describe any entangled states. CHSH inequality $\langle B\rangle|\leq 2$ for any pair of two states among the 3 states is however always satisfied.

\section{Conclusion}

 There exist no non-contextual hidden-variables models in $d=4$ which are fully consistent with quantum mechanics~\cite{beltrametti, gleason, kochen}. Two basic properties missing in the non-contextual hidden-variables representation in (1.1) are the linearity of the probability measure, which is essential to the Born probability interpretation, and the entanglement related to the linearity of Schr\"{o}dinger equations. These two properties are logically independent if one limits the available states, as is seen in the separable quantum mechanical states which are consistent with the linearity of the probability measure but not with entanglement. We have shown that Hardy's model provides a simple test of the consistency of the non-contextual hidden-variables model  in $d=4$ with the linearity of the quantum probability measure. 
 The linearity of the probability measure we analyzed emphasizes the aspects of hidden-variables models which are different from those in the analysis of contextuality~\cite{mermin}. Further analyses along these lines will clarify the physical basis of the important concept of local realism.
\\

\noindent {\bf Acknowledgments}
\\

\noindent We thank Sixia Yu, Qing Chen and other members of Center for Quantum Technologies, National University of Singapore, for very helpful comments. One of the authors (KF) thanks the hospitality of the Center for Quantum Technologies. This work is partially supported by the National Research Foundation and Ministry of Education, Singapore (Grant No. WBS: R-710-000-008-271).

\end{document}